\documentclass{article}
\usepackage{spconf,amsmath,graphicx}

\usepackage{algorithm}
\usepackage{algpseudocode}
\usepackage{fullpage}
\usepackage[utf8]{inputenc} 
\usepackage[T1]{fontenc}    
\usepackage{url}            
\usepackage{booktabs}       
\usepackage{amsfonts}
\usepackage{amsthm}
\usepackage{amsmath,amssymb}
\usepackage{circledsteps}
\newcommand{\vertiii}[1]{{\left\vert\kern-0.25ex\left\vert\kern-0.25ex\left\vert #1 \right\vert\kern-0.25ex\right\vert\kern-0.25ex\right\vert}}

\DeclareMathOperator*{\argmin}{arg\,min}

\usepackage{mathtools}

\theoremstyle{definition}

\theoremstyle{theorem}

\theoremstyle{corollary}

\theoremstyle{prop}

\theoremstyle{lemma}
\newtheorem{lemma}{Lemma}
\theoremstyle{remark}
\newtheorem{remark}{Remark}
\usepackage{nicefrac}       
\usepackage{microtype}      
\usepackage{xcolor}         
\usepackage[export]{adjustbox}
\usepackage{caption}
\usepackage{subcaption}
\usepackage{enumitem}
\usepackage{chngcntr}
\usepackage{apptools}
\newcommand{\transpose}{\mathsf{T}}
\usepackage[ colorlinks = true,
linkcolor = blue,
urlcolor = blue,
citecolor = red,
anchorcolor = green,
]{hyperref}

\title{Differential Analysis for Networks Obeying Conservation Laws}
\name{Anirudh Rayas${}^\ast$, Rajasekhar Anguluri${}^\ast$, Jiajun Cheng${}^\dagger$, Gautam Dasarathy${}^\ast$\thanks{This work was supported in part by the National Science Foundation (NSF) under the grants CCF-2048223, CCF-2007688, and
CBET-2200161, and by the National Institutes of Health (NIH) under the grant 1R01GM140468-01.}}
\address{${}^\ast$School of Electrical, Computer, and Energy Engineering\\
${}^\dagger$School of Computing and Augmented Intelligence\\
Arizona State University, Tempe, AZ, 85281 USA\\
\{ahrayas, rangulur, ccheng58, gautamd\}@asu.edu
}
\begin{document}
%
\maketitle
\begin{abstract}
Networked systems that occur in various domains, such as the power grid, the brain, and opinion networks, are known to obey conservation laws. For instance, electric networks obey Kirchoff's laws, and social networks display opinion consensus. Such conservation laws are often modeled as balance equations that relate appropriate injected flows and potentials at the nodes of the networks. A recent line of work considers the problem of estimating the unknown structure of such networked systems from observations of node potentials (and only the knowledge of the statistics of injected flows). Given the dynamic nature of the systems under consideration, an equally important task is estimating the change in the structure of the network from data -- the so called \emph{differential network analysis} problem. That is, given two sets of node potential observations, the goal is to estimate the structural differences between the underlying networks.  We formulate this novel differential network analysis problem for systems obeying conservation laws and devise a convex estimator to learn the edge changes directly from node potentials. We derive conditions under which the estimate is unique in the high-dimensional regime and devise an efficient ADMM-based approach to perform the estimation. Finally, we demonstrate the performance of our approach on synthetic and benchmark power network data.  

\end{abstract}
\section{Introduction}\label{sec: intro}
A networked system is said to obey a conservation law if flows are neither created nor destroyed. Depending on the context, flows could represent current in electric circuits, water in hydraulic networks, or opinion dynamics in social networks \cite{van2017modeling}. Such systems are at the heart of many natural, engineering, and societal networks \cite{bressan2014flows}. These laws can be conveniently modeled as balance equations that posit a linear map between injected flows and potentials  at the network nodes. For finite-dimensional networks, this linear map is the Laplacian matrix and its sparsity pattern encodes the network structure---the edge connectivity of the network. 

\looseness=-1 In many practical problems of interest, one often does not know the structure of the network, a key information for learning, leveraging, and operating complex systems. Consequently, a recent line of work (see, for example, \cite{anguluri2021grid,rayas2022learning,deka2020graphical,li2020learning,varghese2022transmission}) considers estimating the network structure from observations of node potentials (and only the knowledge of the statistics of injected flows). Given the dynamic nature of the systems under consideration, an equally important task is estimating the change in the structure of network from data. 
This problem, dubbed \emph{differential network analysis}, 
appears in many biological and genomics networks \cite{shojaie2021differential,na2021estimating, yuan2017differential} and is the focus of the paper. 

\looseness=-1 The differential network analysis problem we consider is for systems obeying conservation laws and is stated as follows: given node potential observations from a system at two different time instants, estimate the sparse changes in the network at these time instants. A generalized version of this problem is to estimate the sparse changes in two systems using two sets of node potential observations, one from each system. We distinguish it from the existing differential network analysis in that we exploit the relationship between the injected flows and the potentials.

A na\"ive approach to learning sparse changes is first estimating the individual network structures and then looking for differences in the estimates. Unsurprisingly,  such an indirect approach would be statistically inefficient since it expends effort on estimating parameters that are irrelevant to the task at hand (e.g., edges that do remain unchanged). 
To overcome these issues, we propose an $\ell_1$-norm regularized convex estimator to learn the sparse edge changes directly
using samples from the node potentials. Our estimator exploits the fact that the sparsity pattern of the network is encoded in the \emph{square root} of the inverse covariance matrix of the node potential vector. We derive conditions under which the estimate is unique in the high-dimensional regime. Finally, we present an ADMM approach to numerically solve the estimator and evaluate the performance of our approach on synthetic and benchmark power network data.

\section{Preliminaries and Background}
\label{sec:format}
Let $\mathcal{G}=(V,E)$ be an undirected connected graph on the node set $V \triangleq \{0,1,2,\ldots, p\}$ and edge set $E \subset V\times V$. To each edge $(i,j)$ we associate a non-negative weight $a_{i,j}$. Let $X$ and $Y$ be $p+1$-dimensional real-valued vectors of injections (in-flows) and potentials (out-flows) at the nodes. Then the basic conservation law between these vectors is $X-B^*Y = 0$, 
where $B^*$ is a Laplacian matrix such that $B_{i,j}=-a_{i,j}$ for $i\ne j$ and $B_{i,i}=-\sum_{j=1,i\ne j}a_{i,j}$ for $i=j$.
The key property of $B^*$ is that edge $(i,j)\in E$ if and only if $B_{i,j}\ne 0$.
The model above is sometimes referred to as a generalized Kirchoff's law and is a flexible model describing the relationship between flows and potentials in a variety of systems, including electrical circuits, hydraulic networks, opinion consensus in social networks, etc.; see e.g., \cite{rayas2022learning, van2017modeling} and references therein.    
We work with the reduced graph obtained by deleting the node $0$ and its edges in $\mathcal{G}$. This reduction is standard in many problems (see e.g., \cite{DekaTSG2020, dorfler2012kron}). With an abuse of notation, we denote the Laplacian of the reduced graph as $B^*$. Importantly, $B^*$ is a $p \times p$ positive definite matrix; and hence, invertible \cite{dorfler2012kron}.  The invertiblity assumption ensures that $B^*$ is identifiable from $Y$.

 
\subsection{Differential network analysis}\label{sec: dna}
Consider two networked systems $\mathcal{G}_1$ and $\mathcal{G}_2$ with same node sets but different edge sets. Let $X_1\sim \mathcal{N}(0,\Sigma_{X_1})$ and $X_2\sim \mathcal{N}(0,\Sigma_{X_2})$ be the injection vectors at the nodes of $\mathcal{G}_1$ and $\mathcal{G}_2$. Then, the corresponding node potentials $Y_i = (B_i^{*})^{-1}X_i$ satisfy $Y_i\sim \mathcal{N}(0,{\Theta^{*}}_i^{-1})$ where $\Theta^*_i=B_i^*\Sigma_{X_i}^{-1}B_i^*$ and $i\in \{1,2\}$. We model injections as random vectors to account for unmodelled injections in the system. For example, these injections could be instantaneous consumer demands in power networks. 

\looseness=-1 Recall from Section~\ref{sec: intro} that we are interested in the changes in the network structure. More formally, suppose that we have access to $n_i$ i.i.d samples from $Y_i$. Then, our goal is to estimate $\Delta_{B}^*=B_2^*-B_1^*$. This difference matrix captures changes in the edge weights of $\mathcal{G}_1$ and $\mathcal{G}_2$. Of particular interest is the sparsity pattern of $\Delta_{B}^*$ as it indicates how similar (or dissimilar) the network systems are. 

\looseness=-1 We next develop an expression for $\Delta^*_{B}$ as a function of $\Theta^*_1$ and $\Theta^*_2$ which is a starting point for our algorithm design and analysis. We assume that the injection covariances $\Sigma_{X_i}\succ 0$ are known (see Remark~\ref{rmk: unknown covariance} for relaxing this assumption). We recall an important fact that any positive (semi) definite matrix has a unique square root that is also a positive (semi) definite matrix. That is, for any $C \succeq 0$,  there exists a unique $M\succeq 0$ such that $C=M^2$ \cite{bhatia2009positive}. 

Consider $\Sigma_{X_i}=M_{X_i}^2$ such that $M_{X_i}\succ 0$ is unique. Define $\widetilde{Y}_i=M_{X_i}Y_i$ and set $\widetilde{\Theta}^*_i=(\text{Cov}[\widetilde{Y}_i])^{-1}$. Then
\begin{align}\label{eq: exact_delta}  \Delta^*_{B}=M_{X_2}(\widetilde{\Theta}^*_2)^{\frac{1}{2}}M_{X_2}-M_{X_1}(\widetilde{\Theta}^*_1)^{\frac{1}{2}}M_{X_1}, 
\end{align}
where $(\widetilde{\Theta}^*_i)^{\frac{1}{2}}=M_{X_i}^{-1}{B^*_i}M_{X_i}^{-1}\succ 0$ is the unique square root matrix of $\widetilde{\Theta}^*_i$. 
The expression in \eqref{eq: exact_delta} follows by direct substitution. The uniqueness is because $(\widetilde{\Theta}^*_i)^{-1}=M_{X_i}{\Theta^*}^{-1}_iM_{X_i}=(M_{X_i}{B^*_i}^{-1}M_{X_i})^2\succ 0$.

Because $\Sigma_{X_i}$ are known, their square roots $M_{X_i}$ are known. Therefore, a natural estimator for $\Delta^*_{B}$ is to replace $(\widetilde{\Theta}^*_i)^{\frac{1}{2}}$ in \eqref{eq: exact_delta} with its sample estimate---the square root of the inverse of the sample covariance matrix of $\widetilde{Y}_i$. This estimate is unfortunately highly sample inefficient. In fact, the sample covariance matrix is non-invertible when $p>n_{i}$ (the so-called high-dimensional regime). Alternatively, we can estimate $(\widetilde{\Theta}^*_i)^{\frac{1}{2}}$ using well-known estimators such as GLASSO or CLIME \cite{yuan2007model,friedman2008sparse,cai2011constrained}. But these estimators work well only when $\Theta_i^*$ is sparse. As shown in our prior work \cite{rayas2022learning}, $\Theta_i^*$ need not be sparse even when $B_i^*$ is sparse. 



We overcome the challenges above by directly estimating $\Delta^*_B$ assuming it is sparse. This assumption is mild compared to the stringent assumption that $B_i^*$ is sparse and it is satisfied in several applications (see Introduction). 

\begin{remark}\textit{(Unknown covariance matrix $\Sigma_{X_i}$.)} If $\Sigma_{X_i}$ is unknown, we can slightly modify $\Delta^*_B$ to estimate differences between the square roots $(\Theta^*_1)^{\frac{1}{2}}$ and $(\Theta^*_2)^{\frac{1}{2}}$ (given in \eqref{eq: exact_delta}). This approach works best if the sparsity of $B_i^\ast$ (approximately) equals the sparsity of $(\Theta^*_1)^{\frac{1}{2}}$, which for instance happens when $\Sigma_{X_i}$ is (approximately) diagonal. \label{rmk: unknown covariance}
\end{remark}


\section{A convex square root estimator}
We introduce our square root difference estimator to estimate a sparse $\Delta^*_B$. Let 
$\widetilde{\Psi}_i=M^{-1}_{X_i}(\widetilde{\Theta}^{*}_i)^{-\frac{1}{2}}M^{-1}_{X_i}$, where $(\widetilde{\Theta}^{*}_i)^{-\frac{1}{2}}=(\text{Cov}[\widetilde{Y}_i])^{\frac{1}{2}}$ as defined in \eqref{eq: exact_delta}. Let $\Delta \in \mathbb{R}^{p\times p}$ and consider the following loss function: 
{\small 
\begin{align}\label{eq: D-trace loss function}
    \mathcal{L}(\Delta)&=\frac{1}{4}(\langle\widetilde{\Psi}_1\Delta, \Delta\widetilde{\Psi}_2\rangle+\langle\widetilde{\Psi}_2\Delta, \Delta\widetilde{\Psi}_1\rangle)
    -\langle\Delta, \widetilde{\Psi}_1-\widetilde{\Psi}_2\rangle,
\end{align}
}%
where $\langle A, B\rangle \triangleq \mathrm{tr}(AB^\transpose)$. Such loss functions, dubbed D-trace losses, have emerged as a computationally efficient alternative to the log-det loss function and are related to score-matching losses~\cite{zhang2014sparse}. 
Understanding statistical properties of estimators based on D-trace loss functions is an active study of research \cite{yuan2017differential,wu2020differential,xudong2017confidence}. 

A loss-function similar to \eqref{eq: D-trace loss function} has been used in \cite{yuan2017differential} to learn the difference between two graphical models using the covariance matrices. Instead, we learn the difference between two networks using the square roots of the covariance matrices. Using the arguments in \cite{yuan2017differential}, we can show that the loss in \eqref{eq: D-trace loss function} is convex in $\Delta$. If $\Sigma_i \succ 0$, the unique minima for this loss function occurs at  $\Delta^*_B=B_2^*-B_1^*$. So to obtain a sparse estimate of $\Delta^*_B$ using the samples of $Y_i$, we solve the $\ell_1$-regularized optimization problem: 
\begin{align}\label{eq: convex estimator}
    \widehat{\Delta}_B\in \argmin_{\Delta \in \mathbb{R}^{p\times p}} \mathcal{L}(\Delta)+\lambda_n\|\Delta\|_{1,\text{off}}, 
\end{align}
where $\lambda_n\!\geq\!0$ and $\Vert \Delta\Vert_{1,\text{off}}\!=\!\sum_{i\neq j}\vert \Delta_{ij}\vert$ is the $\ell_1$-norm applied on the off-diagonal elements of $\Delta$. The estimate
 $\widehat{\Psi}_i=M_{X_i}^{-1}\widetilde{S}_i^{\frac{1}{2}}M_{X_i}^{-1}$, where ${\widetilde{S}_i}^\frac{1}{2}$ is the unique square root of the sample covariance matrix of $\widetilde{Y}_i$ and $i\in \{1,2\}$. 

In Section \ref{sec: opt} we develop an iterative procedure to solve \eqref{eq: convex estimator}. We conclude this section by stating a result on the uniqueness of $\widehat{\Delta}_B$ in \eqref{eq: convex estimator}. Because $\|\cdot\|_1$ is convex, the combined loss function in \eqref{eq: convex estimator} is strongly convex provided the Hessian of $\mathcal{L}(\cdot)$ is positive definite. Then, we can invoke KKT conditions for strongly convex functions to conclude that $\widehat{\Delta}_B$ is unique. However, unfortunately, the Hessian matrix $ H\triangleq (\widehat{\Psi}_1\otimes \widehat{\Psi}_2+\widehat{\Psi}_2\otimes \widehat{\Psi}_1)/2$ is only positive semi definite. This is because $\widehat{\Psi}_i$ is positive semi definite when $p>n_i$. Hence, $\widehat{\Delta}_B$ is not unique.

Nonetheless, Lemma \ref{lma: uniq soln} below establishes the uniqueness of $\widehat{\Delta}_B$ by placing certain restrictions on the nullspace of the Hessian matrix. Lemma \ref{lma: uniq soln} is in the spirit of uniqueness results in compressed sensing. Let $\text{vec}(A)$ be the $mn$-dimensional vector obtained by stacking the columns of $A \in \mathbb{R}^{m\times n}$ on top of each other. Let $\text{vec}^{-1}(z)$, for any $z\in \mathbb{R}^{mn}$, be such that $\text{vec}^{-1}(\text{vec}(A))=A$. Define the nullspace or kernel of $A$ as $\mathrm{Ker}(A)=\{d: Ad=0\}$. 

\begin{lemma}\label{lma: uniq soln} Let $H$ be defined as above and $d \in \mathrm{Ker}(H)$. Then $\widehat{\Delta}_B$ in \eqref{eq: convex estimator} is unique if and only if $d^\transpose\text{vec}(\widehat{\Psi}_1-\widehat{\Psi}_2)\leq 0$ and $\|\text{vec}^{-1}(d)\|_{1,\text{off}}\leq \tau$. Here, $\tau$ is given by the constrained (Lagrangian dual) form of \eqref{eq: convex estimator} and $d\ne 0$. 
\end{lemma}
We sketch a few details of the proof here. Consider the constrained form of \eqref{eq: convex estimator}: $\argmin_{\|\Delta\|_{1,\text{off}}\leq \tau} \mathcal{L}(\Delta,\widehat{\Psi}_1,\widehat{\Psi}_2)$. Now, rewriting the loss function in \eqref{eq: D-trace loss function} in its quadratic form we have: $\text{vec}(\Delta)^\transpose H \text{vec}(\Delta)/2-\text{vec}(\Delta)^\transpose(\widehat{\Psi}_1-\widehat{\Psi}_2)$. The uniqueness result then follows from  \cite{dostal2009solvability}.

\section{Optimization algorithm}\label{sec: opt}
We solve the optimization in \eqref{eq: convex estimator} using the alternating direction method of multipliers (ADMM) method proposed in \cite{yuan2017differential}.
We give high level details of this method while referring the reader to \cite{yuan2017differential} for complete details.

Consider the following identity for the loss function in \eqref{eq: D-trace loss function}: 
$\mathcal{L}(\Delta,\widehat{\Psi}_1,\widehat{\Psi}_2)=(\mathcal{L}_1(\Delta))/4+(\mathcal{L}_2(\Delta))/4$, where $    \mathcal{L}_1(\Delta)=\langle\widehat{\Psi}_1\Delta^*_B, \Delta^*_B\widehat{\Psi}_2\rangle-2\langle \Delta, \widehat{\Psi}_1-\widehat{\Psi}_2\rangle$,
and similarly, $\mathcal{L}_2(\Delta)=\langle\widehat{\Psi}_2\Delta^*_B, \Delta^*_B\widehat{\Psi}_1\rangle-2\langle \Delta, \widehat{\Psi}_1-\widehat{\Psi}_2\rangle$. The only change in these loss functions is the positioning of $\widehat{\Psi}_1$ and $\widehat{\Psi}_2$. Consider three $p\times p$ matrices $\Delta_1$, $\Delta_2$, and $\Delta_3$. Then, the optimization in \eqref{eq: convex estimator} is equivalent to
\begin{align}\label{eq: auxilary}
    \argmin_{\Delta_1\!=\!\Delta_2\!=\!\Delta_3}\hspace{-1.0mm} \mathcal{L}_1(\Delta_1)\!+\!\mathcal{L}_2(\Delta_2)\!+\!\lambda_n\|\Delta_3\|_1, 
\end{align}
where $\widehat{\Delta}_B=\widehat{\Delta}_i$ for any $i\in \{1,2,3\}$. Let $\rho >0$ be the momentum constant, and $\Lambda_1$, $\Lambda_2$, and $\Lambda_3$ be the matrix multipliers of the augmented Lagrangian of \eqref{eq: auxilary} (see \cite{yuan2017differential} for a formula). Then, the ADMM iterates are
\begin{align*}
    \Delta_1^{k+1} &= G(\widehat{\Psi}_1,\widehat{\Psi}_2,2\rho\Delta_3^k+2\rho\Delta_2^k+\widehat{\Psi}_1-\widehat{\Psi}_2+\\
   & \quad \quad 2\Lambda_1^k-2\Lambda_3^k,4\rho),\\
  \Delta_2^{k+1} &= G(\widehat{\Psi}_2,\widehat{\Psi}_1,2\rho\Delta_3^k+2\rho\Delta_1^{k+1}+\widehat{\Psi}_1-\widehat{\Psi}_2+\\
  & \quad \quad 2\Lambda_{3}^k-2\Lambda_2^k,4\rho),\\
  \Delta_3^{k+1} &= S((\rho\Delta_1^{k+1}+\rho\Delta_2^{k+1}\!-\!\Lambda_1^{k}+\Lambda_2^k)/(2\rho),\lambda/2\rho),\\
\Lambda_1^{k+1} &= \Lambda_1^k+\rho(\Delta_3^{k+1}-\Delta_1^{k+1}),\\
\Lambda_2^{k+1} &= \Lambda_2^k+\rho(\Delta_2^{k+1}-\Delta_3^{k+1}), \text{and}\\
\Lambda_3^{k+1} &= \Lambda_3^k+\rho(\Delta_1^{k+1}-\Delta_2^{k+1}). 
\end{align*}
The shrink function $S(\cdot)$ is defined as follows: $S(A, \lambda) = 0$ when $\left| A \right| \leq \lambda$, and $S(A, \lambda) = A - {\rm sign}(A)\lambda$.
For any symmetric matrices $P$, $Q$, and $R$, and a positive $\gamma$, the function $G(\cdot)$ takes the following form: $G(P,Q,R,\gamma)$ $\triangleq U_P\{O\circ(U_P^\transpose R U_Q)\}U_Q^\transpose$, where $\circ$ is the Hadamard product of two matrices. Further, $U_P{D}_PU_P^\transpose$ and $U_Q{D}_QU_Q^\transpose$ are the eigendecompositions of $P$ and $Q$. Finally, $O_{ij}=[D_{P}(j,j)D_{Q}(i,i)+\gamma]^{-1}$. The formula for $G(\cdot)$ in \cite{yuan2017differential} is incorrect and the expression we state here is correct.

\section{Numerical Simulations}
\label{sec:typestyle}

We illustrate the performance of our estimator on synthetic and two benchmark power systems. We consider two performance metrics: (i) the empirical probability (averaged over 100 instances) of recovering the support of $\Delta^*_B$ and (ii) the worst case error evaluated using $\|\widehat{\Delta}_{B}-\Delta^{\ast}_B\|_{\infty}$. Recall that $\|A\|_\infty=\max_{i,j}|a_{i,j}|$. In the figures below, we plot these error metrics as a function of the re-scaled sample size $n/(d^2\log(p))$, where $d$ is the maximum degree of $\Delta^{\ast}_{B}$. This scaling is theoretically justified in \cite{yuan2017differential}. We set $\lambda_{n}\propto\sqrt{\log(p)/n}$ and the parameter $\rho = 0.001$.  
\vspace{-2.0mm}

\begin{figure}[h!]
    \centering
    \includegraphics[width=1.0\linewidth]{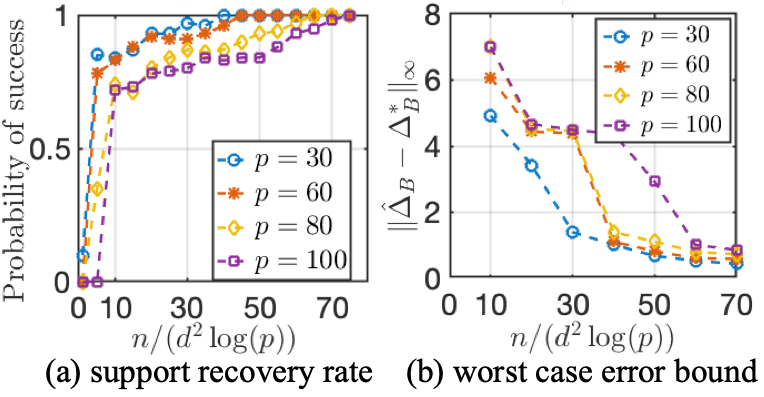}
    \caption{Estimation accuracy for synthetic networks.}
    \label{fig:Grid graph plots}
\end{figure}

Figure.~\ref{fig:Grid graph plots} shows the estimation accuracy for $\Delta_B^*$ for many dimensions $(p)$. In each case, the graph underlying $\Delta^*_B$ is a grid graph with degree $d=4$. We can visualize this graph by letting the nodes correspond to the points in the 2D-plane with integer coordinates. For this choice of $\Delta^*_B$, we set $B_1^*$ to be a random, invertible symmetric matrix. We then define $B_2^*=B_1^*+\Delta_B^*$. Importantly, $B_2^*$ and $B_1^*$ are non-sparse. In Fig.~\ref{fig:Grid graph plots}(a) and (b), the accuracy improves as a function of the re-scaled sample size. But the accuracy deteriorates as the dimension ($p$) increases, which is expected. Notably, for fixed $d$, $\|\widehat{\Delta}_{B} - \Delta^{\ast}_{B}\|_{\infty}$ behaves approximately as $1/\sqrt{n/(\log(p))}$, which agrees with the support recovery results on sparse regression. 

\vspace{-2.0mm}
\begin{figure}[h!]
    \centering
    \includegraphics[width=1.0\linewidth]{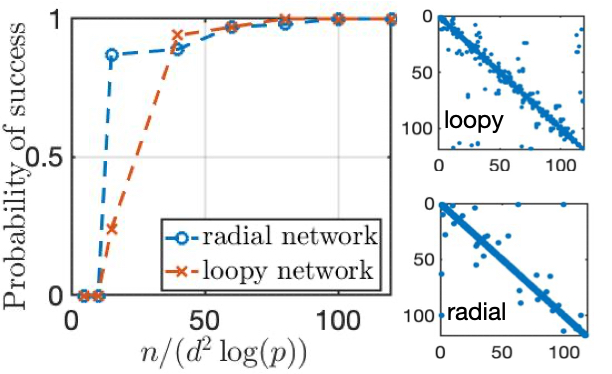}
    \caption{Estimation accuracy for power networks.}
    \label{fig:benchmark_plot}
\end{figure}

Similar to Figure.~\ref{fig:Grid graph plots}, Figure.~\ref{fig:benchmark_plot}, shows the estimation accuracy for different choices of $\Delta^*_B$ whose underlying graphs are grids. But $B_1^*$ and the graph underlying it are associated with an electric power network. Specifically, we consider the radial IEEE 118 bus distribution network and the loopy IEEE 118 bus transmission network \cite{zimmerman2005matpower}. As mentioned earlier, we reduced the networks by deleting a node. Hence, $p=117$. The panels on the right visualize the sparsity patterns of the reduced networks. For both networks, the support recovery rate in the left panel increases with the re-scaled sample size. This result again confirms that the sparsity of individual networks plays no role in the estimation performance. 

\vspace{-2.0mm}
\begin{figure}[h!]
    \centering
    \includegraphics[width=0.90\linewidth]{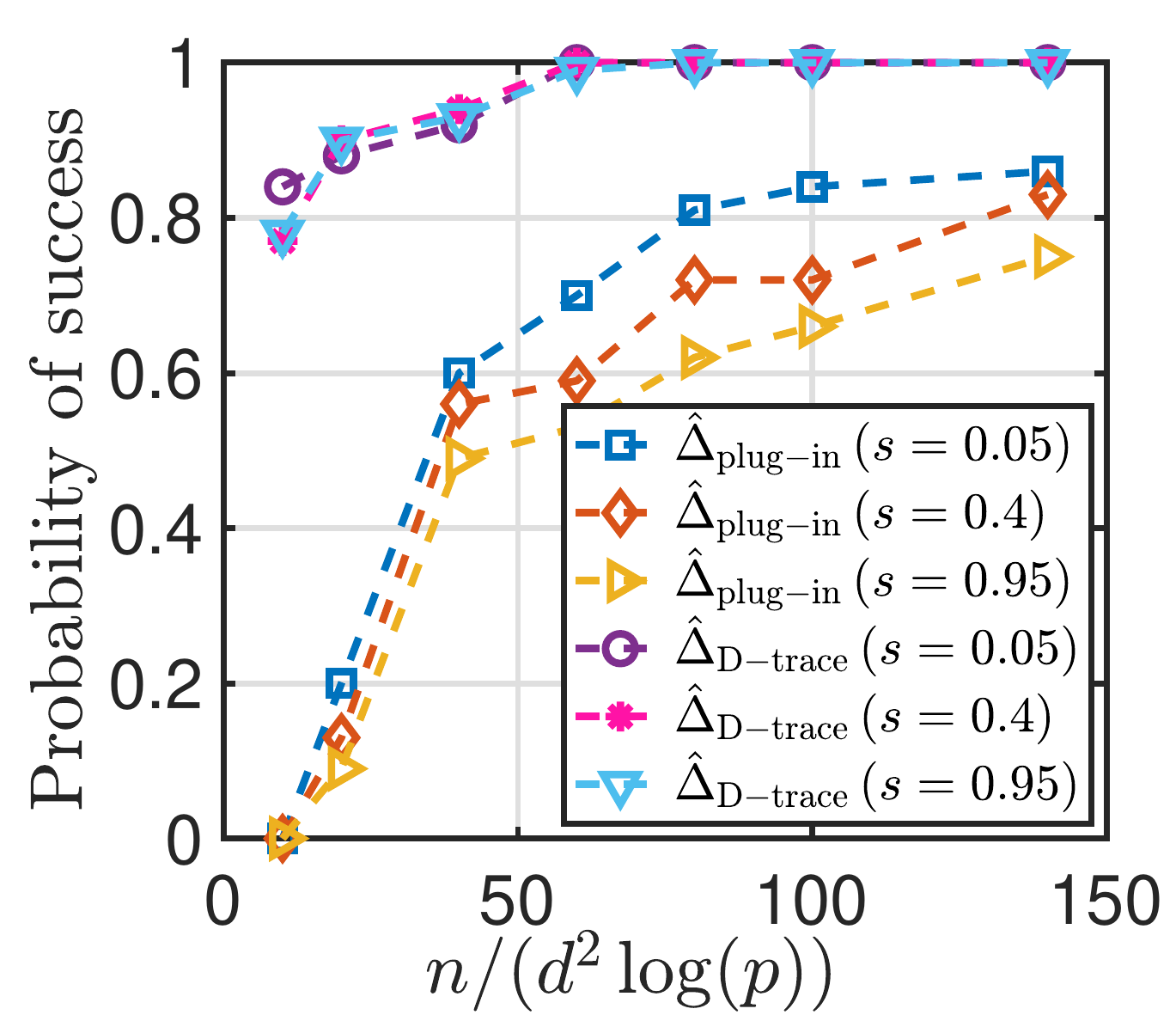}
    \caption{Plug-in estimator vs proposed estimator.}
    \label{fig:Comparison with plug-in}
\end{figure}

Figure.~\ref{fig:Comparison with plug-in} compares the support recovery rates of the proposed estimator and the naive plug-in estimator. The latter is obtained by plugging the inverse of the square root of the sample covariance matrix in \eqref{eq: exact_delta}.  So, for this experiment, we assume that $n>p=60$. We consider three matrices for $B_1^*$, with increasing number of zeros. We regulate the number of zeros in $B_1^*$ using the parameter $s$, which is defined as the ratio of the number of non-zeros to the number of entries in the matrix. The smaller the $s$, the sparser is the matrix. The graph underlying $\Delta^*_B$ is grid and we let $B_2^*=B_1^*+\Delta^*_B$. As shown in Figure.~\ref{fig:Comparison with plug-in}, for every choice of $s$, our estimator (called D-trace in the figure), outperforms the plug-in estimator. Importantly, our estimator works well even when $n<p=60$, where the plug-in estimator does not even exist.

\section{Conclusion}
\label{sec:majhead}

\looseness=-1 In this paper, we consider differential network analysis for systems obeying conservation laws. For random node injections, we show that the sparsity pattern of the square root of the inverse covariance matrix of the node potential vector encodes the network structure. We exploit this property to develop an estimator that directly estimates the difference of two network Laplacian matrices using the samples of potentials. We adapt the ADMM method in \cite{yuan2017differential} to numerically implement the proposed estimator. Our numerical results demonstrate the superior performance of our estimator over the standard plug-in estimator.  


 




\bibliographystyle{IEEEtran}
\bibliography{strings,refs}

\end{document}